\documentclass[conference]{IEEEtran}
\IEEEoverridecommandlockouts

\usepackage{cite}
\usepackage{bm}
\usepackage{amsmath,amssymb,amsfonts}
\usepackage{algorithmic}
\usepackage{graphicx}
\usepackage{textcomp}
\usepackage{xcolor}
\usepackage{multirow}
\usepackage{tabularx}
\usepackage{enumitem}
\usepackage{algorithm}
\usepackage{amsmath}
\usepackage{amssymb}
\usepackage{algorithmic} 
\usepackage{amsthm}
\usepackage{url}
\usepackage{textcomp}
\usepackage{xcolor}
\usepackage{array} 
\usepackage{booktabs,makecell, multirow, tabularx}
\usepackage{cite}
\usepackage{bm}
\usepackage{amsmath,amssymb,amsfonts}
\usepackage{algorithmic}
\usepackage{graphicx}
\usepackage{textcomp}

\usepackage{array} 
\usepackage{booktabs,makecell, multirow, tabularx}

\newcommand{\sysname}{HYDG}
\def\BibTeX{{\rm B\kern-.05em{\sc i\kern-.025em b}\kern-.08em
    T\kern-.1667em\lower.7ex\hbox{E}\kern-.125emX}}
\begin{document}

\title{Hypergraph-Based Dynamic Graph Node Classification}
\author{
\IEEEauthorblockN{Xiaoxu Ma$^1$, Chen Zhao$^2$, Minglai Shao$^1$\textsuperscript{*}\thanks{*Corresponding author.}, Yujie Lin$^1$}
\IEEEauthorblockA{
$^1$\textit{School of New Media and Communication, Tianjin University, Tianjin, China} \\
$^2$\textit{Department of Computer Science, Baylor University, Waco, Texas, USA} \\
\{maxiaoxu, shaoml, linyujie\_22\}@tju.edu.cn, chen\_zhao@baylor.edu}}

\maketitle

\begin{abstract}
Node classification on static graphs has achieved significant success, but achieving accurate node classification on dynamic graphs where node topology, attributes, and labels change over time has not been well addressed. Existing methods based on RNNs and self-attention only aggregate features of the same node across different time slices, which cannot adequately address and capture the diverse dynamic changes in dynamic graphs. Therefore, we propose a novel model named Hypergraph-Based Multi-granularity Dynamic Graph Node Classification (\textbf{\sysname{}}). After obtaining basic node representations for each slice through a GNN backbone, \sysname{} models the representations of each node in the dynamic graph through two modules. The individual-level hypergraph captures the spatio-temporal node representations between individual nodes, while the group-level hypergraph captures the multi-granularity group temporal representations among nodes of the same class. Each hyperedge captures different temporal dependencies of varying lengths by connecting multiple nodes within specific time ranges. More accurate representations are obtained through weighted information propagation and aggregation by the hypergraph neural network. Extensive experiments on five real dynamic graph datasets using two GNN backbones demonstrate the superiority of our proposed framework.
\end{abstract}
\begin{IEEEkeywords}
Dynamic Graph, Hypergraph, Node Classification
\end{IEEEkeywords}
\section{Introduction}
    \label{sec:intro}

As a widely employed data structure, graph structures excel in representing intricate data and possess robust capabilities for storing and articulating node attributes and entity relationships. The current landscape of graph neural networks\cite{gcn,GAT,tian2024mldgg,wang2024madod,he2024gdda,shao2024supervised,tian2024graphs} predominantly operates within the realm of static graph structures\cite{graphsage}. However, many real-world scenarios involve dynamic graph structures, such as recommender systems, social media analytics\cite{liu2019characterizing}, transportation systems\cite{li2023dynamic}, citation networks\cite{zhang2019dane}and epidemic modeling. Dynamic graphs can be broadly classified into discrete-time dynamic graphs and continuous-time dynamic graphs\cite{skarding2021foundations}.In both forms of dynamic graphs, node features, node labels, and graph structures may evolve over time\cite{dysat}, posing a challenge for traditional static graph models in capturing the dynamic relationships among node features. Consequently, the effective capture and utilization of node relationships and features in dynamic networks have emerged as pivotal research questions.
\begin{figure}[]
  \centering
  \setlength{\abovecaptionskip}{0.cm}
  \includegraphics[width=0.47\textwidth,scale=1.00]{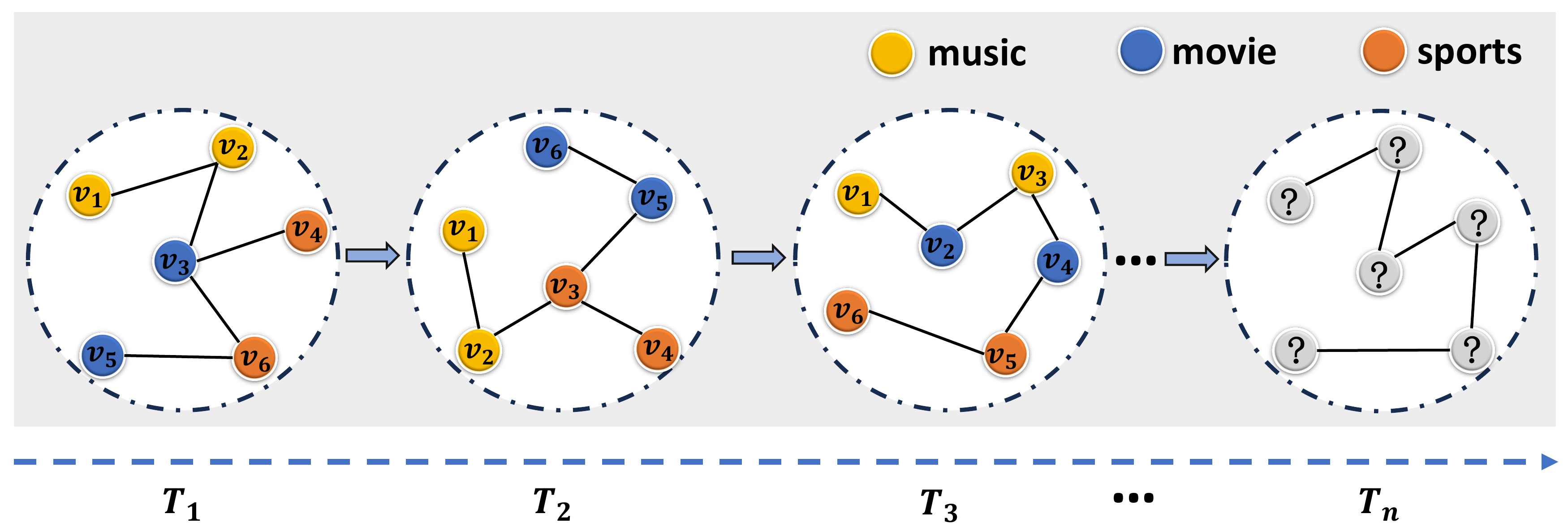}
  \caption{An illustrative example of interest prediction in dynamic social graphs involves predicting users' interest tags at a specific time based on historical data. In this network, both user connections and interest tags evolve over time, and our goal is to forecast users' interest tags using information from the interest network.}
  \vspace{-20pt}
  \label{fig:intro}
\end{figure}




In this paper, we focus on the task of node classification in discrete dynamic graphs, specifically predicting the node labels in unknown future graph snapshots given the graph information at previous time steps. As illustrated in Fig.~ \ref{fig:intro}, in social interest networks, a user’s social connections and personal information may change significantly over time, leading to shifts in their interest labels. Current methods primarily fall into two categories\cite{skarding2021foundations,barros2021survey}: one uses RNNs\cite{manessi2020dynamic,evolvegcn,GCNLSTM,RNNGCN} to capture temporal relationships between nodes, while the other employs attention\cite{GCNSE,dysat} mechanisms to capture dynamic relationships between nodes. However, in diverse real-world dynamic networks, simply using RNNs or self-attention to link the representations of the same node across different time slices may fail to adequately capture the features of various node categories. In scenarios involving changes in node features, label shifts, node disappearance, or the addition of new nodes, traditional dynamic graph learning methods struggle to perform inductive learning effectively, thus failing to cope with the complex node dynamics within dynamic networks.

To address these challenges, we propose the Hypergraph-Based Multi-granularity Dynamic Graph Node Classification (\textbf{\sysname{}}) algorithm, which captures both individual and group-level node features using hypergraphs\cite{zhou2006learning,gao2022hgnn+,gao2020hypergraphsurvey,yan2020learning}. By connecting multiple nodes within specific time intervals through hyperedges, \sysname{} constructs multi-granularity features, enabling more accurate classifications via hypergraph neural networks. Compared to traditional methods, \sysname{} more effectively models high-order dependencies and enhances temporal information capture at both individual and group levels through weighted information propagation. Our contributions include:
\begin{itemize}[leftmargin=*]
\item We introduce a novel hypergraph-based algorithm for dynamic graph node classification that effectively captures diverse and robust spatio-temporal dependencies through hypergraph construction, weighted node propagation, and feature aggregation.
\item We develop a multi-scale hypergraph construction approach that concurrently captures both individual- and group-level node features, enhancing the diversity of node representations.
\item Our model, built upon two GNN backbones, demonstrates superior performance on five real-world datasets, consistently outperforming baseline models.
\end{itemize}
\section{methodology}
    \label{sec:method}
    The primary focus of this paper is to address the problem of node classification on discrete dynamic graphs. Given the dynamic graph snapshots $\textbf{G}_{P} = \{\textbf{G}^1, \textbf{G}^2, \ldots, \textbf{G}^t\}$ up to time step $t$ and the corresponding node labels $\textbf{Y}_{P} = \{\textbf{Y}^1, \textbf{Y}^2, \ldots, \textbf{Y}^t\}$, our goal is to classify all nodes in future graph snapshots $\textbf{G}_{F} = \{\textbf{G}^{t+1}, \textbf{G}^{t+2}, \ldots, \textbf{G}^T\}$, whose labels $\textbf{Y}_{F} = \{\textbf{Y}^{t+1}, \textbf{Y}^{t+2}, \ldots, \textbf{Y}^T\}$ are unknown. 
\begin{figure*}[t]
  \centering
  \setlength{\abovecaptionskip}{0.cm}
  \includegraphics[width=0.85\textwidth]{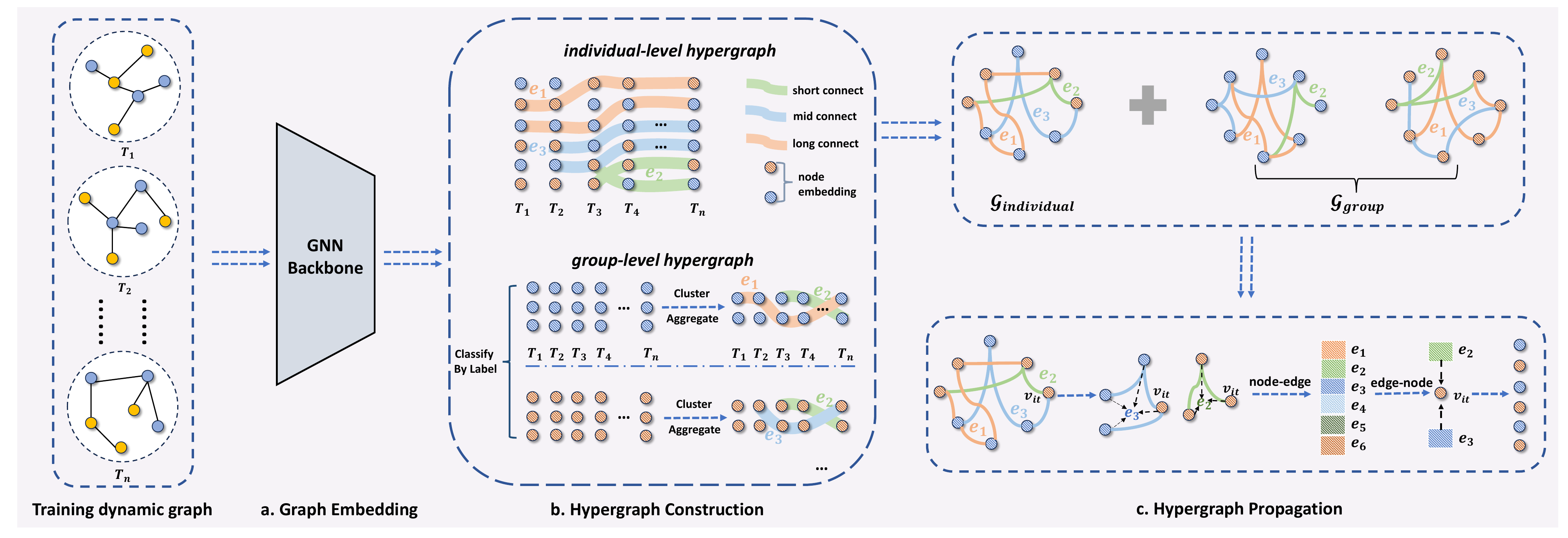}
  \caption{ The detailed architecture of \sysname{} consists of three components: graph embedding(blue and orange
represent different labels), hypergraph construction (for better visualization, we only illustrate the binary clustering approach),and hypergraph propagation.}
  \vspace{-10pt}
  \label{fig:model}
\end{figure*}
\subsection{Dynamic Graph Feature Extraction}
Static graph neural networks have been widely applied in graph representation learning tasks.  Therefore, we first utilize a backbone GNN to perform feature extraction on each node of every temporal slice, as shown in Fig.~\ref{fig:model}(a). This process yields a feature representation for each node at each time step.
For the t-th temporal slice of the dynamic graph $\textbf{G}^t$, we can utilize it to obtain feature representations.
\begin{equation}\label{eq:feature-extraction}
\mathbf{Z}^t = \text{GNN}(\mathbf{A}^t, \mathbf{X}^t),
\end{equation}where $\mathbf{Z}^t$ represents the matrix of feature embeddings for the nodes in the graph snapshot $\mathbf{G}^t$ processed by the GNN, with $\mathbf{Z}^t = [\mathbf{z}_{1}^t, \mathbf{z}_{2}^t, \ldots, \mathbf{z}_{n}^t]$, where $n$ is the number of nodes. $\mathbf{A}^t$ denotes the adjacency matrix of $\mathbf{G}^t$, and $\mathbf{X}^t$ refers to the initial feature matrix of the nodes in $\mathbf{G}^t$.

\subsection{Hypergraph Construction}
In dynamic graphs, nodes possess both individual features and common features shared within their subgroups. To better capture the higher-order correlations of nodes within each time slice, we construct multi-scale hypergraphs by capturing  individual-level and group-level  dependencies, as shown in Fig.~\ref{fig:model}(b).

\textbf{Individual-level hypergraph construction.}~ We capture the individual high-order dependencies of each node in the dynamic graph by constructing a series of individual hyperedges, denoted as $\mathcal{G}_{in} = (\mathcal{V}_{in}, \mathcal{E}_{in})$. Here, $\mathcal{G}_{in}$ includes individual nodes $\mathcal{V}_{in}$ and the set of hyperedges $\mathcal{E}_{in}$. Specifically, for a given node \(v_{it}\), we identify \(K\) nodes from other time slices within a specified time range that are most close to the feature vector obtained from  (\ref{eq:feature-extraction}), excluding the current time slice. These \(K+1\) nodes are then connected via a hyperedge $\mathbf{e}_{it}$, as shown in (\ref{eq:eKNN}).
\begin{equation}\label{eq:eKNN}
     \mathbf{e}_{it} = \{v_{it}, \forall v_{jt'}\in \mathcal{N}_K(v_{it})\}, s.t. \ |t - t'| \leq \tau~, 
\end{equation}    
where $\mathcal{N}_K$ represents the set of $K$ nearest neighbors, which can be calculated using various distance metrics such as Euclidean distance, Chebyshev distance, cosine distance, etc. $|*|$ denotes the temporal distance between nodes. $\tau$ represents the threshold value of the time range. We model the multi-scale temporal dependencies in the dynamic graph by setting three threshold ranges: short-term connection, mid-term connection, and long-term connection, respectively. We construct the individual hypergraph $\mathcal{G}_{in}$ using $v_{it}$ and $e_{it}$ obtained from all nodes in each time slice.




\textbf{Group-level hypergraph construction.~}To better capture the multi-granularity group features of each category in dynamic graphs, we construct a series of group-level hypergraphs $\mathcal{G}_{group}$, where $\mathcal{G}_{group} = (\mathcal{V}_{group},\mathcal{E}_{group})$, including group nodes $\mathcal{V}_{group}$ and group hyperedge sets $\mathcal{E}_{group}$. 
In summary, we first group the feature vectors $\mathbf{Z}^t$ according to the true labels $\mathbf{Y}^t$ of the nodes, followed by hierarchical clustering and aggregation of these groups within each time slice. This process yields spatio-temporal group representations, which are then used to construct hypergraphs for each node category, capturing both the features and their spatio-temporal dependencies in the dynamic network.
The mathematical representation is as follows:

\vspace{-10pt}
 \begin{equation}\label{eq:aggregate}
\mathbf{Z}^{c,t}_{group} = \{Agg\left(Cluster_M(\{z_v^t \mid y_i^t = c\})\right)\},
\end{equation} 
where $y_v^t$ denote the label of node $v$ at time $t$. By grouping node embeddings with the same label, we obtain $\mathbf{O}^{c,t} = \{ \mathbf{z}_v^t \mid y_v^t = c \}$, where $C$ represents the number of node categories. The function $Cluster_M(\cdot)$ performs clustering on the grouped embeddings. Applying $Cluster_M(\mathbf{O}^{c,t})$ yields the clusters $\{ \mathbf{O}^{c,t}_1, \mathbf{O}^{c,t}_2, \ldots, \mathbf{O}^{c,t}_M \}$, where $M$ denotes the number of clusters. The aggregation function $Agg(\cdot)$ is applied to aggregate the vectors within each cluster, which could be operations such as $\max(\cdot)$, $\min(\cdot)$, or $avg(\cdot)$. Finally, for each group, we obtain $\mathbf{Z}^{c,T}_{group} \in \mathbb{R}^{M \times T}$. This process is formalized to capture diverse temporal characteristics of nodes within the same category across different clusters at various time steps.

We merge each time slice obtained from $\mathbf{Z}^{c,t}_{group}$ by (\ref{eq:aggregate}) to obtain the grouped $\mathbf{Z}^{c,T}_{group}$. Following the construction method of individual-level hypergraph in (\ref{eq:eKNN}), we construct hyperedges for $\mathbf{Z}^{c,T}_{group}$ separately to capture the spatio-temporal evolutionary relationships of various category group characteristics, resulting in $\mathcal{V}_{group}^c$ and $\mathcal{E}_{group}^c$.
Hence, we obtain $\mathcal{G}_{group}=\{\mathcal{G}_{group}^c\}_{c \in C}$.








\subsection{Hypergraph Propagation}

After constructing individual-level hypergraph $\mathcal{G}_{in}$ and group-level hypergraph $\mathcal{G}_{group}$, We employ Hypergraph Neural Networks (HGNN)\cite{feng2019hypergraph} for node-weighted information propagation and feature updating to obtain spatio-temporal fusion features within hyperedges, as illustrated in Fig.~\ref{fig:model}(c).

For a given hypergraph $\mathcal{G}$, the incidence matrix $\mathbf{H}$ is defined based on the node set $\mathcal{V}$ and edge set $\mathcal{E}$ as:
\begin{equation}\label{eq:H}
    H(v, e) = 1 \text{ if } v \in e, \text{ else } 0,\quad \forall v \in \mathcal{V}, \forall e \in \mathcal{E}.
\end{equation}

Based on $\mathcal{G}_{in}$ and $\mathcal{G}_{group}$, we can derive the individual and group-level incidence matrices, $\mathbf{H}_{in}$ and $\mathbf{H}_{group}$, through (\ref{eq:H}).
 Following the approach of frequency-domain GCNs \cite{gcn}, we use Fourier transformation to shift spatial signals to the frequency domain for convolution, and then apply the inverse Fourier transformation \cite{feng2019hypergraph}, as formalized by the following equation:
 \begin{equation}\label{eq:HGNN}
\mathbf{Z}^{(l+1)} = \sigma \left( \mathbf{D}_v^{-\frac{1}{2}} \mathbf{H} \mathbf{W} \mathbf{D}_e^{-1} \mathbf{H}^T \mathbf{D}_v^{-\frac{1}{2}} \mathbf{Z}^{(l)} \Theta^{(l)} \right),
\end{equation}where $\mathbf{Z}$ represents the feature matrix, $\mathbf{H}$ denotes the incidence matrix, $\Theta$ is the learnable hypergraph convolution kernel, $\mathbf{W}$ is the learnable weight matrix, and $\sigma$ is the non-linear activation function.
More specifically, for a given node \( v_{it} \), \( H(v_{it}) \) represents all hyperedges connected to node \(  v_{it} \), containing nodes with the highest relevance to \( v_{it}\). We aggregate the features of nodes from different time slices in the hyperedges, except for \(  v_{it} \), using the corresponding weights \( \mathbf{w}_{it}^{jt'} \) in \(  H(v_{it}) \) to capture the temporal dependencies within the same hyperedge:
\begin{equation}\label{eq:node-weight}
    \mathbf{w_{it}^{jt'}}= \exp \left(-\frac{d(\mathbf{z}_{it}, \mathbf{z}_{jt'})^2}{\sigma^2}\right),\forall v_{jt'}\in \mathbf{e}_{it},
\end{equation}    
\vspace{-8pt} 
\begin{equation}\label{eq:node-edge}
    \mathbf{p}^{k,l}_{it} = \sum_{j \neq i}^{v_{jt'} \in e_{k}} \mathbf{z}^{l-1}_{jt'}\mathbf{w_{it}^{jt'}}, \ \forall  \mathbf{e}_k \in H(v_{it}),
\end{equation}
$v_{it}$ and $v_{jt'}$ represent a pair of nodes at different time slices within a hyperedge. $d(\mathbf{z}_{it}, \mathbf{z}_{jt'})$ denotes the distance between $v_{it}$ and $v_{jt'}$ . We can use various methods to compute the distance, such as Euclidean distance, cosine distance, etc. $\sigma$ can be set as the median of distances between all pairs of vertices, $\mathbf{z}^{l-1}_{jt'}$ represents the output of node $v_{jt'}$ at layer $l-1$ in HGNN.

After obtaining different hyperedge features for the dynamic graph node \( v_{it} \), we compute the weight by measuring the similarity between the hyperedge features and node features and normalize the weights using softmax function. Then, we perform edge-node level spatio-temporal dependency aggregation:
\begin{equation}\label{eq:combined}
\mathbf{z}^{l}_{it} = \sum_{k} \frac{{\rm exp}( sim(\mathbf{z}^{k,l-1}_{it}, \mathbf{p}^{k,l}_{it}))}{\sum_{k'}{\rm exp}(sim(\mathbf{z}^{k',l-1}_{it}, \mathbf{p}^{k',l}_{it}))} \mathbf{p}^{k,l}_{it}.
\end{equation}

The node representations obtained from the above process, denoted as $\mathbf{z}^{l}_{it}$ ,are passed through a non-linear function, yielding the integrated representation of node dependencies across time and space.

\subsection{Model Learning}

After the hypergraph information propagation, we obtain the individual and group hypergraph representation, $\mathbf{Z}_{in}$ and $\mathbf{Z}_{group}$.By applying the softmax function, for each time slice, we compute the predicted labels $\mathbf{\hat{Y}}_{in}$ and $\mathbf{\hat{Y}}_{group}$ for the nodes in $\mathcal{G}_{in}$ and $\mathcal{G}_{group}$, respectively. The true labels for the group-level hypergraph, $\mathbf{Z}_{group}^c$, are represented by $c$ as follows:
\begin{equation}\label{eq:individual-level loss}
    \mathcal{L}_{in}(\mathbf{Y}, \mathbf{\hat{Y}}) = - \sum_{i=1}^N y_i \log( \frac{e^{z_{in}^i}}{\sum_{j=1}^{N} e^{z_{in}^j}})~,
\end{equation}
\vspace{-5pt} 
 \begin{equation}\label{eq:group-level loss}
   \mathcal{L}_{group}(\mathbf{Y},\mathbf{\hat{Y}}) =  -\sum_{c=1}^C\sum_{i=1}^M c \log( \frac{e^{z_{group}^{ic}}}{\sum_{j=1}^{N} e^{z_{group}^{ic}}})~.
\end{equation}   

For $\mathcal{L}_{in}$ and $\mathcal{L}_{group}$, we assign different weights, denoted as $\alpha$ and $\beta$, respectively, resulting in the overall loss function $\mathcal{L}_{all}$, which consists of the following two terms:
\begin{equation}\label{eq:loss all}
     \mathcal{L}_{all} = \alpha \cdot  \mathcal{L}_{in} + \beta \cdot  \mathcal{L}_{group}~.
\end{equation}

During training, we construct both individual and group-level hypergraphs to capture diverse node representations using hypergraph neural networks. During testing, without labels, we use only individual-level hypergraphs and the trained network for predictions. The group-level hypergraphs, used for data augmentation, capture evolving relationships within the same class, addressing variations in unseen test sets.

\section{experiments}
    \label{sec:exp}
    \begin{table*}[htbp]
    \centering
    \small
    \caption{Quantitative results on the dynamic node classification task. The results are the averages of five runs, with the best result in each column highlighted in bold and the second best result indicated with a dash.}
    \label{tab:baselines}
    \resizebox{\textwidth}{!}{
    \begin{tabular}{c|l|c c| c c| c c| c c| c c}
        \toprule
        \textbf{backbone}&\textbf{Dataset} & \multicolumn{2}{c|}{\texttt{DBLP5}} & \multicolumn{2}{c|}{\texttt{DBLP3}} & \multicolumn{2}{c|}
        {\texttt{Reddit}} & \multicolumn{2}{c|}{\texttt{Wiki}} & \multicolumn{2}{c}{\texttt{ML-RATING}} \\
        &\textbf{Metric} & \textbf{Accuracy} & \textbf{AUC} & \textbf{Accuracy} & \textbf{AUC} & \textbf{Accuracy} & \textbf{AUC} & \textbf{Accuracy} & \textbf{AUC} & \textbf{Accuracy} & \textbf{AUC} \\
        \midrule
        &GCN\cite{gcn}&$63.01\pm0.62$ &$0.65\pm0.02$&$74.23\pm0.41$&$0.55\pm0.02$&$31.02\pm0.33$&$\underline{0.52\pm0.01}$&$79.22\pm0.81$&$0.65\pm0.01$&$85.53\pm0.22$&$0.67\pm0.01$\\
         &GCNLSTM\cite{GCNLSTM} &$61.83\pm1.84$ &$0.65\pm0.01$&$73.37\pm0.92$&$0.56\pm0.01$&$30.26\pm0.42$&$0.50\pm0.01$&$79.43\pm1.22$&$\textbf{0.68}\pm\textbf{0.01}$&$84.77\pm0.84$&$\underline{0.68\pm0.02}$ \\
       &RNNGCN\cite{RNNGCN} &$65.92\pm0.43$ &$\textbf{0.70}\pm\textbf{0.01}$&$76.55\pm0.41$&$0.56\pm0.01$&$\underline{32.68\pm0.28}$&$0.51\pm0.01$&$78.74\pm1.18$&$0.61\pm0.01$&$\underline{85.75\pm0.52}$&$0.66\pm0.01$ \\
       &GCNSE\cite{GCNSE} &$66.58\pm0.31$ &$0.64\pm0.02$&$\underline{76.88\pm0.52}$&$0.53\pm0.01$&$31.09\pm0.47$&$0.51\pm0.01$&$79.49\pm0.62$&$0.54\pm0.01$&$85.63\pm0.34$&$0.57\pm0.01$ \\
         \textbf{GCN}&ROLAND\cite{roland}  &$\underline{66.63\pm0.38}$&$0.67\pm0.00$&$76.15\pm0.42$&$0.56\pm0.01$&$32.31\pm0.39$&$\underline{0.52\pm0.01}$&$\underline{79.63\pm0.34}$&$0.59\pm0.04$&$85.69\pm0.28$&$\textbf{0.70}\pm\textbf{0.01}$ \\
          &EvolveGCN\cite{evolvegcn}  &$63.21\pm1.24$ &$0.57\pm0.02$&$75.18\pm0.62$&$0.54\pm0.02$&$26.32\pm0.65$&$0.50\pm0.01$&$76.25\pm1.01$&$0.57\pm0.03$&$81.46\pm0.36$&$0.56\pm 0.06$ \\
        &DySAT\cite{dysat}  &$65.75\pm0.81$ &$0.62\pm0.02$&$75.36\pm0.44$&$\underline{0.57\pm0.01}$&$29.39\pm0.73$&$0.51\pm0.01$&$77.36\pm0.92$&$0.60\pm0.01$&$85.19\pm0.61$&$0.65\pm0.02$ \\
        \cmidrule{2-12}
        &\textbf{\sysname{}-GCN} (ours) &$\textbf{68.26}\pm\textbf{0.20}$&$\underline{0.69\pm0.00}$&$\textbf{77.12}\pm\textbf{0.13}$&$\textbf{0.58}\pm\textbf{0.01}$&$\textbf{34.12}\pm\textbf{0.42}$&$\textbf{0.53}\pm\textbf{0.01}$&$\textbf{80.67}\pm\textbf{0.26}$&\underline{$0.67\pm0.02$}&$\textbf{86.96}\pm\textbf{0.17}$&$\textbf{0.70}\pm\textbf{0.01}$ \\
        \midrule
          &GraphSAGE\cite{graphsage} &$67.83\pm0.69$ &\underline{$0.75\pm0.01$}&$74.06\pm0.31$&\underline{$0.61\pm0.00$}&$31.85\pm0.85$&$0.51\pm0.00$&$77.83\pm1.32$&$0.63\pm0.05$&$84.15\pm1.45$&$0.68\pm0.02$ \\
         &GCNLSTM\cite{GCNLSTM} &$67.32\pm1.25$ &$0.64\pm0.01$&$75.75\pm0.58$&$0.58\pm0.01$&$32.03\pm0.53$&$\textbf{0.54}\pm\textbf{0.00}$&$\underline{78.36\pm1.19}$&$0.62\pm0.03$&$83.76\pm2.24$&$0.69\pm0.02$ \\
        &RNNGCN\cite{RNNGCN} &$65.31\pm0.83$ &$0.74\pm0.01$&\underline{$76.66\pm0.72$}&$0.56\pm0.01$&$31.53\pm0.98$&$0.52\pm0.01$&$77.30\pm1.04$&$\textbf{0.67}\pm\textbf{0.01}$&$\underline{85.65\pm1.17}$&$0.69\pm0.02$ \\
        &GCNSE\cite{GCNSE} &$67.36\pm0.97$ &$0.52\pm0.02$&$\textbf{76.94}\pm\textbf{0.57}$&$0.53\pm0.01$&$31.02\pm0.69$&$0.50\pm0.01$&$78.06\pm1.16$&$0.60\pm0.01$&$85.29\pm1.32$&$0.62\pm0.03$ \\
        \textbf{GraphSAGE}&ROlAND\cite{roland} &\underline{$68.42\pm0.52$} &$0.74\pm0.01$&$75.68\pm0.83$&$0.60\pm0.01$&\underline{$32.61\pm0.43$}&\underline{$0.53\pm0.00$}&$78.22\pm0.59$&$0.64\pm0.00$&$85.23\pm0.76$&\underline{$0.70\pm0.01$} \\
        &EvolveGCN\cite{evolvegcn}  &$63.21\pm1.24$ &$0.57\pm0.02$&$75.18\pm0.62$&$0.54\pm0.02$&$26.32\pm0.65$&$0.50\pm0.01$&$76.25\pm1.01$&$0.57\pm0.03$&$82.46\pm0.36$&$0.57\pm 0.06$ \\
        &DySAT\cite{dysat}  &$65.75\pm0.81$ &$0.65\pm0.02$&$75.36\pm0.44$&$0.57\pm0.01$&$29.39\pm0.73$&$0.51\pm0.01$&$77.36\pm0.92$&$0.60\pm0.01$&$85.19\pm0.61$&$0.67\pm0.01$ \\
        \cmidrule{2-12}
        &\textbf{\sysname{}-SAGE} (ours) &$\textbf{70.40}\pm\textbf{0.32}$&$\textbf{0.77}\pm\textbf{0.00}$&$76.51\pm0.13$&$\textbf{0.63}\pm\textbf{0.01}$&$\textbf{33.98}\pm\textbf{0.24}$&$\textbf{0.54}\pm\textbf{0.00}$&$\textbf{80.53}\pm\textbf{0.47}$&\underline{$0.66\pm0.01$}&$\textbf{86.95}\pm\textbf{0.35}$&$\textbf{0.72}\pm\textbf{0.01}$ \\
        \bottomrule
    \end{tabular}
    }
    \vspace{-5pt}  
\end{table*}
\subsection{Experimental Settings}
We conducted experiments using five real-world dynamic graph datasets, with detailed information provided in Table \ref{tab:datasets}, We selected seven baselines for comparison, including both static\cite{gcn,graphsage} and dynamic graph neural networks\cite{dysat,RNNGCN,GCNSE,evolvegcn,roland}. The primary task is to predict the node labels in the test set at various time slices using the node features and label information from limited time slices in the training set. To better evaluate the model's performance, we used accuracy and Macro-AUC as evaluation metrics. Each method employs a two-layer graph neural network, utilizing both GCN and GraphSAGE as GNN layers for experimentation.
\vspace{-10pt}
\begin{table}[h]
    \centering
    \small
    \caption{Real datasets for evaluation of our methods.}
    \label{tab:dataset}
     \resizebox{0.47\textwidth}{!}{
    \begin{tabular}{cccccc}
        \hline
        Dataset & Nodes & Edges & Time Steps & Classes & Attributes \\
        \hline
        \texttt{DBLP3}\cite{RNNGCN} & 4257 & 23540 & 10 & 3 & 100 \\
        \texttt{DBLP5}\cite{RNNGCN} & 6606 & 42815 & 10 & 5 & 100 \\
        \texttt{Reddit}\cite{GCNSE} & 8291 & 264050 & 10 & 4 & 20 \\
        \texttt{Wiki}\cite{wiki_dataset} & 9227 & 31910 & 11 & 4 & - \\
        \texttt{ML-Rating}\cite{harper2015movielens} & 9746 & 2000438 & 11 & 7 & - \\
        \hline
    \end{tabular}}
    \vspace{-10pt}
    
    \label{tab:datasets}
\end{table}

\subsection{Overall Performance}


As shown in Table \ref{tab:baselines}, we conduct experiments using both GCN and GraphSAGE backbones, with EvolveGCN and DySAT sharing experimental results across both sets. The results demonstrate that the \sysname{} model consistently outperforms baseline models in node classification tasks across five datasets. By constructing individual hypergraphs and multi-scale group-level hypergraphs while capturing spatio-temporal dependencies, \sysname{} achieves superior accuracy on the \texttt{DBLP5}, \texttt{Reddit}, \texttt{Wiki}, and \texttt{ML-RATING} datasets using both GCN and GraphSAGE backbones, and records the highest AUC scores on the \texttt{DBLP3}, \texttt{Reddit}, and \texttt{ML-RATING} datasets. Notably, \sysname{} shows a 2\%-3\% improvement in accuracy on the \texttt{DBLP5} and \texttt{Reddit} datasets compared to the best-performing baselines, despite slightly trailing RNN-GCN in AUC on \texttt{DBLP3}. These results suggest that the ability of \sysname{} to capture spatio-temporal dependencies enables better learning of node representations and relationships in dynamic graphs, outperforming traditional methods based on self-attention mechanisms and RNNs.

\subsection{Ablation Study}
\begin{figure}[htbp]
  \centering
  \setlength{\abovecaptionskip}{0.cm}
  \includegraphics[width=0.49\textwidth,scale=1.00]{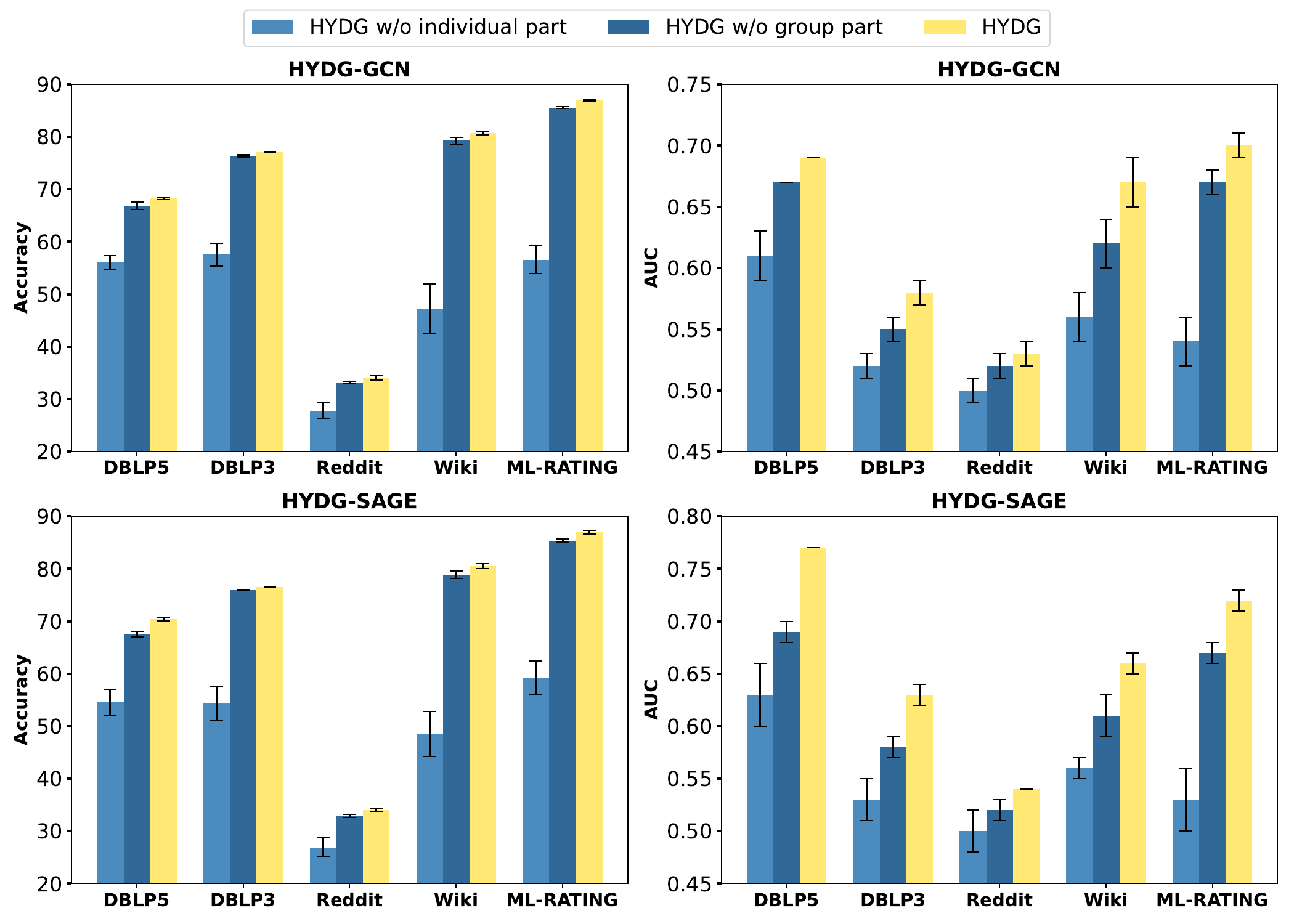}
  \caption{Ablation study results of \sysname{} on node classification tasks across five datasets.}
  \vspace{-5pt}
  \label{fig:ablation1}
\end{figure}

To better explore the roles of individual-level hypergraphs and group-level hypergraphs in capturing spatio-temporal dependencies in dynamic graphs, we conduct ablation experiments on these two modules separately, and the results are shown in Fig.~\ref{fig:ablation1}. \textbf{(i) Removal of individual-level hypergraphs. }Utilizing only group-level hypergraphs yields poor accuracy and AUC results. This is because group-level hypergraphs capture features only within each slice, resulting in insufficient usable nodes and inadequate capture of features and dependencies across samples. \textbf{(ii) Removal of group-level hypergraphs. }Utilizing only individual-level hypergraphs achieves relatively high accuracy but results in lower AUC. 
This is because the individual hypergraph construction, which links hyperedges by identifying each node's nearest neighbors, may capture only single dynamic dependency patterns. Additionally, connections between nodes with different labels can lead to incomplete dynamic feature representations.
\vspace{-20pt}

\section{conclusion}
    \label{sec:conclusion}
    This paper presents a dynamic graph node classification framework based on multi-granularity hypergraphs. By constructing individual-level hypergraphs across various time ranges and multi-granularity group-level hypergraphs, the framework effectively captures higher-order spatio-temporal dependencies in dynamic graph nodes. Hypergraph neural networks are employed for weighted information propagation, leading to more accurate and robust node representations. Extensive experiments on five real datasets using two backbones demonstrate that the proposed framework outperforms the optimal baseline models.
\vfill\pagebreak

\bibliographystyle{IEEEtran}
\bibliography{IEEEabrv,refs}
\end{document}